\documentclass{ws-procs9x6}

\begin{document}

\title{Inhibition of Transport of a Bose-Einstein Condensate in a Random Potential}

\author{J.~A.~Retter$^{1}$, A.~F.~Var\'{o}n$^{1}$, D.~Cl\'{e}ment$^{1}$, M.~Hugbart$^{1}$, P.~Bouyer$^{1}$, L.~Sanchez-Palencia$^{1}$,  D.~Gangardt$^{2}$, G.~V.~Shlyapnikov$^{2,3}$, and~A.~Aspect$^{1}$}

\address{$^{1}$Laboratoire Charles Fabry de l'Institut d'Optique,\\
Universit\'e~Paris-Sud~XI, 91403~Orsay cedex, France.\\
$^{2}$Laboratoire de Physique Th\'{e}orique et Mod\`{e}les
Statistiques, Universit\'e~Paris-Sud~XI, 91405~Orsay cedex,
France. \\$^{3}$Van der Waals-Zeeman Institute, University of
Amsterdam, Valckenierstraat~65/67, 1018 XE Amsterdam, The
Netherlands.}

\maketitle

\abstracts{We observe the suppression of the 1D transport of an
interacting elongated Bose-Einstein condensate in a random
potential with a standard deviation small compared to the typical
energy per atom, dominated by the interaction energy. Numerical
solutions of the Gross-Pitaevskii equation reproduce well our
observations. We propose a scenario for disorder-induced trapping
of the condensate in agreement with our
observations\cite{Clement}.}

\section{Introduction}

Coherent transport of waves in disordered systems is a topic of
primary importance in condensed-matter physics, for example in the
description of normal metallic conduction, superconductivity and
superfluid flow, and has relevance also to optics and
acoustics\cite{Akkermans}. The presence of disorder can lead to
intriguing and non-intuitive phenomena such as Anderson
localization\cite{anderson}, percolation dynamics\cite{aharony},
and disorder-driven quantum phase transitions to
Bose-glass\cite{fisher} or spin-glass\cite{parisi} phases.  The
main difficulty in understanding quantum transport arises from the
subtle interplay of scattering on the potential landscape,
interferences and interparticle interactions.  Due to a high
degree of control and measurement possibilities, dilute atomic
Bose-Einstein condensates in optical potentials have proved an
ideal system in which to revisit traditional condensed matter
problems, and recent theoretical works discuss disorder-induced
phenomena in this context\cite{damski2003}. Apart from the
(undesired) effect of a rough potential on trapped cold atoms on
atom chips\cite{chips}, there have been few experiments on BEC in
random potentials\cite{Lye,Arlt}.

In this experiment\cite{Clement}, we study the axial expansion of
an elongated BEC in a magnetic guide, in the presence of a
disordered potential generated by laser speckle. We observe that a
weak disordered potential inhibits both the expansion and the
centre-of-mass (COM) motion of the condensate.

\section{Laser Speckle --- a disordered potential for atoms}
Laser speckle\cite{goodman} is the random intensity pattern
produced when coherent laser light is scattered from a rough
surface. Such patterns arise from interferences between wavefronts
coming from different scattering sites.  In the Fraunhofer limit,
a speckle pattern does not depend on the details of the scattering
surface, but follows a well-defined statistical distribution.  The
intensity distribution is exponential $P(I)=\exp{(-I/\sigma_I)}$
with the intensity standard deviation equal to the local mean
intensity: $\sigma_I=\langle I(z) \rangle$. The typical distance
$\Delta z$ between speckle `grains' can be characterised by the
half-width of the autocorrelation function. For a circular
aperture, this is an Airy function, with the first zero located at
$\Delta z = 1.22 \lambda l / D$, where $\lambda$ is the laser
wavelength, $l$ the distance from the scattering plate to the
observation plane, and $D$ the aperture diameter at the scattering
surface.

\begin{figure}
\begin{center}
\includegraphics[width=11cm]{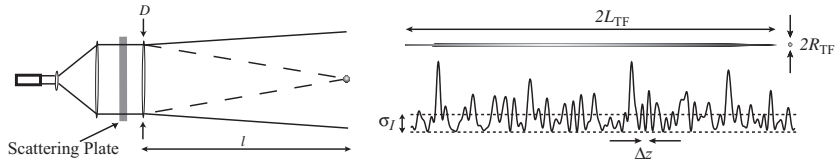}
\end{center}
\caption{Left: Optical setup used to create the random speckle
potential. The condensate is at the focal point of the lens
system, with its long axis oriented perpendicular to the page.
Right: Example of speckle intensity profile, with condensate to
scale. The speckle potential is effectively 1D for the trapped
condensate.}\label{setup}
\end{figure}

In our experiment, a blue-detuned laser beam is shone onto the
atoms through a scattering plate, as shown in Fig.~\ref{setup}.
The beam is derived from a tapered amplifier, injected by a
free-running diode laser at $\lambda\sim 780$\,nm and
fibre-coupled to the experiment. The out-coupled beam is expanded
and then focused onto the condensate, the fibre out-coupler and
lenses being mounted on a single small optical bench, aligned
perpendicular to the long axis of the BEC. Inserting the
scattering plate in the position shown projects an optical speckle
potential onto the atoms, with an intensity distribution
$I(\bf{r})$ which is simply the Fourier transform of the phase
distribution at the scattering plate. The scattered beam has a
total power of up to $150$\,mW and diverges to an rms radius of
1.83\,mm at the condensate. The mean intensity (the gaussian
envelope) of the beam can be assumed constant over the region
where the atoms are trapped.

To calibrate the speckle pattern, the optical set-up is removed
from the BEC apparatus and the intensity distribution observed on
a CCD camera at the same distance $l$ as the atoms. Taking images
with various beam apertures $D$, we verify the exponential
intensity distribution $P(I)$, and compute the autocorrelation
function to obtain the grain size $\Delta z$. Taking into account
the modulation transfer function of the camera\cite{Hug05}, we
find that the measured grain size follows the prediction. For our
setup, $l=140(5)\,$mm and $D=25.4(1)\,$mm, giving $\Delta z=
5.2(2)~\mu$m.  In this experiment, we produce condensates with an
aspect ratio of 100,  typical Thomas-Fermi half-length
$L_{\rm{TF}}=150\,\mu$m and radius $R_{\rm{TF}}=1.5\,\mu$m.  The
trapped BEC occupies about 45--50 minima along its length, but
experiences an almost constant potential in the radial directions.
(Along the axis of the laser beam, the typical length scale of the
speckle grains is much longer, $\sim\Delta z^2/\lambda =
35\,\mu$m.)  The speckle potential is therefore effectively
one-dimensional (1D) for the atoms: $R_{\rm{TF}} < \Delta z \ll
L_\textrm{TF}$. This is important as systems in 1D potentials are
known to be more sensitive to the effects of
disorder\cite{Thouless}.

We calibrate the speckle intensity $\langle I(z) \rangle$ as a
function of the fibre-coupled laser power, from which the
light-shift potential $V(z) \propto I(z)  / \delta$ is
determined\cite{Clement}.  The longitudinal positioning of the
optical set-up at a distance $l$ from the atoms is controlled to
within 5\,mm, leading to an uncertainty of $3.5\%$ on the speckle
intensity. Since the light is blue-detuned from resonance
($|\delta| > 0.15$\,nm), the potential is {\it repulsive} for the
atoms. For the laser intensities used in this experiment, the
standard deviation of the optical potential $\sigma_V=\langle V(z)
\rangle$ is always smaller than the chemical potential $\mu$ of
the condensate.

Other recent experiments have also generated disordered optical
potentials for BEC: using red-detuned laser speckle\cite{Lye} and
by imaging randomly patterned structures onto an optical
lattice\cite{Arlt}.

\section{Experimental Sequence}

\quad\ {\textbf{Condensate preparation.}} We generate an elongated
(quasi-1D) BEC of $^{87}$Rb atoms in the 5S$_{1/2}|F=1,
m_{F}=-1\rangle$ state, using an iron-core electromagnet
Ioffe-Pritchard trap\cite{orsayBEC} with final trap frequencies of
$\omega_{\bot}=2\pi \times 660(4)\,$Hz radially and
$\omega_{z}=2\pi \times 6.70(7)\,$Hz axially. We obtain
condensates of typically $3.5\times 10^5$ atoms and chemical
potential $\mu/2\pi\hbar\sim 5\,$kHz.  In such an elongated
 trapping geometry, phase fluctuations can be
important\cite{Pet01}. However, at our estimated temperature of
$150\,$nK, the phase coherence length is of the order of the
condensate length.

The laser speckle is turned on at the end of the rf evaporation
ramp and we wait a further 200\,ms in the presence of the rf field
to ensure that the condensate is in equilibrium at the end of the
sequence.

{\textbf{Axial expansion.}}  We next open the magnetic trap in the
axial (dipole) direction while keeping the transverse confinement
and the random potential unchanged, thus converting the trap into
a long, uniform, magnetic guide. The condensate expands along this
guide, the axial expansion being driven by the repulsive
interatomic interactions.

Due to hysteresis in the electromagnet and coupling between the
dipole and quadrupole poles, remanent magnetizations can produce
an axial trapping potential even when the dipole excitation
current is reduced to zero. For this reason it is necessary to
invert the current in the dipole excitation coils. To avoid
spin-flip losses to non-trapped hyperfine states, the minimum
magnetic field must not cross zero, thus setting a lower limit for
the dipole current. By extrapolating the variation of the
condensate dipole mode frequency with the current, we estimate the
final axial trapping frequency to be less than $\omega_{z} \sim
2\pi \times 0.4$\,Hz, which is compatible with the linear
expansion observed (see Fig.~\ref{Fig.2}).  The trap opening is
carried out over 30\,ms, in order to minimize trap loss and
heating. Numerical simulations of the 3D Gross-Pitaevskii (GP)
equation confirm that this `slow' relaxation of the axial
confinement reduces only slightly (by $\sim 4\,\%$) the asymptotic
expansion velocity of the BEC  with respect to a `sudden' trap
opening.

{\textbf{Imaging.}} After an axial expansion time $\tau$, 
the magnetic fields and laser speckle are switched off and the
condensate is imaged by absorption after a further 15\,ms
time-of-flight. From axial profiles of these images, the COM
position and rms half-length $L$ of the atomic cloud are
determined.

\section{Inhibition of Transport}
\begin{figure}
\begin{center}
\includegraphics[width=11cm]{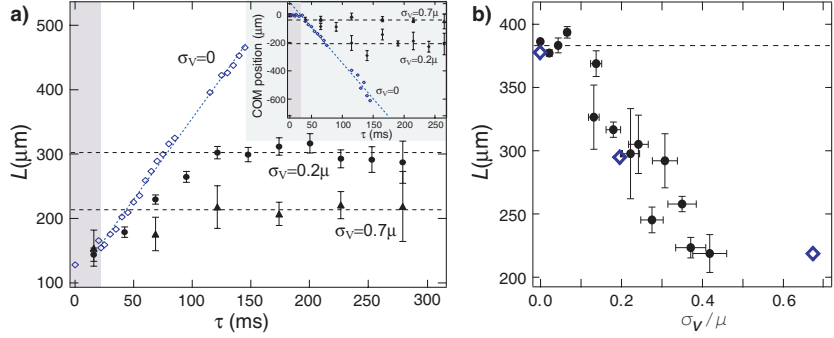}
\end{center}
\caption{{\textbf{a)}} Rms half-length $L$ and COM position
(inset) of the BEC as a function of axial expansion time $\tau$,
for three different amplitudes of the disordered potential:
$\sigma_V=0$, $0.2\mu$, $0.7\mu$. The axial confinement is relaxed
during the first 30\,ms (grey band). {\textbf{b)}} Variation of
rms half-length $L$ with speckle potential $\sigma_V$ for
$\tau=115\,$ms. The three points $\diamondsuit$ correspond to the
data shown in {\textbf{a)}}.}\label{Fig.2}
\end{figure}

In Fig.~\ref{Fig.2}a we plot the half-length $L$ and COM position
of the condensate as a function of axial expansion time $\tau$. In
the absence of the speckle potential, the condensate expands
axially, reaching a linear expansion velocity of
$2.47(3)\,$mm\,s$^{-1}$, consistent with the numerical predictions
of the GP equation. When a small speckle potential $\sigma_V=0.2
\gamma$ is applied, the expansion is decelerated and the BEC
eventually stops expanding. This effect becomes more pronounced at
higher speckle potentials, as shown in Fig.~\ref{Fig.2}b. The BEC
also acquires a constant COM velocity of 5.1(2)\,mm\,s$^{-1}$ due
to a magnetic `kick' during the trap-opening.  This motion too is
decelerated and finally stopped in the presence of a weak speckle
potential.

By displacing the speckle potential by small distances in the
vertical (radial) direction, we can project a completely different
realization of the speckle potential onto the atoms. Since the BEC
in the magnetic trap already covers many peaks of the speckle
potential, we find that the expansion dynamics are fairly
insensitive to the exact random potential.

\section{`Disorder-induced trapping'}
\begin{figure}[b!]
\begin{center}
\includegraphics[width=11cm]{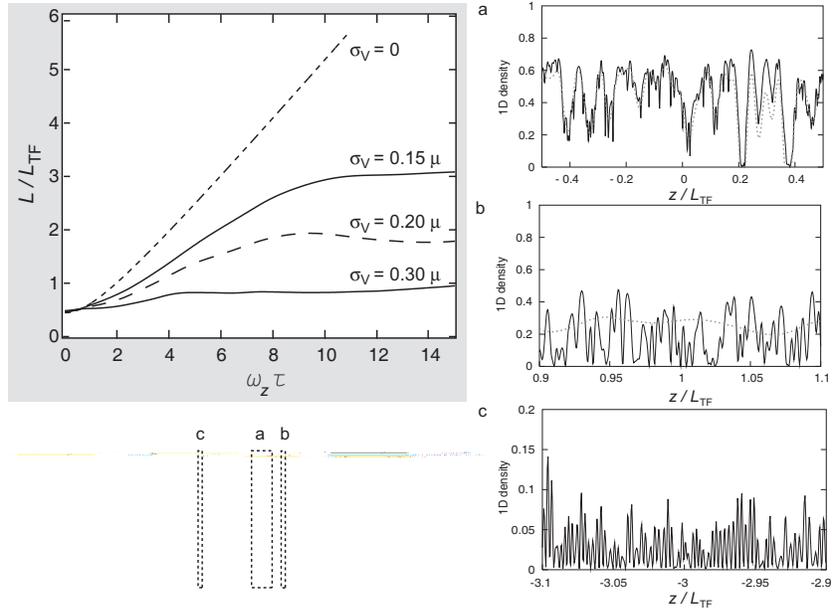}
\end{center}
\caption{\textbf{Top} Time evolution of the rms size $L$ of the
BEC in the random potential, for various speckle amplitudes
$\sigma_V$, as obtained by numerical integration of the 1D GP
equation. {\textbf{Bottom}} Simulated density profile (solid) at
time $\omega_z \tau = 10$ during axial expansion, with
$\sigma_V=0.2\mu$ (grey line represents $V(z)/g_{1D}$).
{\textbf{Right}} Zoom on density profile (solid) --- note the
different length scales. The approximation
$n(z)={\rm{max}}(0,\overline{\mu}(t) - V(z))/g_{1D}$ (dashed)
reproduces the density only in the central region (a) (see
text).}\label{Fig.3}
\end{figure}

We are able to interpret the observed behaviour by studying the
density profiles obtained by numerical integration of the GP
equation. By assuming tight radial confinement, that is $\hbar
\omega_\bot \gg \hbar \omega_z, \mu, k_\textrm{B}T$, this equation
is simplified to 1D. The static speckle potential is generated
numerically, with $\Delta z = 0.049 L_{\rm{TF}}$. In
Fig.~\ref{Fig.3} we see that this model reproduces well the trends
observed experimentally, despite the approximation to 1D.

In the harmonic trap, the BEC is in the interaction-dominated,
Thomas-Fermi regime: $n(z,t=0)={\rm{max}}(0, \mu - 0.5 m
\omega_{z}^2 z^2 - V(z))/g_{1D}$. Since the BEC healing length
$\xi$ is much smaller than $\Delta z$, the density profile simply
follows the modulations of the speckle potential.  These
modulations are not sufficiently strong to fragment the BEC, as
$\sigma_V < \mu$.
After switching off the axial confinement, the BEC expands (for
$\sigma_V=0$) according to  the scaling theory\cite{scaling}. For
times $\omega_z \tau \gtrsim 1$ the fastest atoms populate the
wings of the BEC, where the density is lowest; conversely, at the
centre of the BEC, the density and therefore the interaction
energy remain relatively high, and the kinetic energy very low. In
the presence of disorder we can therefore expect very different
behaviour in each of these two regions.

Figure~\ref{Fig.3} shows a simulated density profile obtained for
$\sigma_V=0.2 \mu$ and $\omega_z \tau=10$, at which time the BEC
has stopped expanding.  Looking more closely at the central region
$-L_{\rm{TF}}/2<z<L_{\rm{TF}}/2$ (Fig.~\ref{Fig.3}a) we find that
the condensate density modulations have mainly the same
length-scale as the speckle potential, and that these modulations
are stationary. Since in this region the interaction energy
remains dominant, we can suppose that the Thomas-Fermi
approximation still holds. By defining an effective chemical
potential $\overline{\mu}(t)$, which decreases with the density
during expansion, we can write the density profile as:
$n(z,t)={\textrm{max}}(0,\overline{\mu}(t) - V(z))/g_{1D}$. This
is a good approximation close to $z=0$ (see Fig.~\ref{Fig.3}a),
showing that the local density adiabatically adapts to the speckle
potential as $\overline{\mu}(t)$ decreases. In this central
region, the expansion stops when the condensate encounters two
large peaks of the speckle potential with amplitudes larger than
$\overline{\mu}(t)$, and the condensate then fragments.

In the low density regions, for example $-3.1
L_{\rm{TF}}<z<-2.9L_{\rm{TF}}$ (Fig.~\ref{Fig.3}c), the density
profile is not stationary, and furthermore the length-scale of the
density modulations is much smaller than $\Delta z$. By using
Fourier analysis, we identify the characteristic length-scale as
the healing length of the trapped condensate,
$\xi=\hbar/\sqrt{2m\mu}$. This can be understood by conservation
of energy, implying that the kinetic energy per atom is of the
order of the typical energy per atom $\mu$, of the initial BEC. In
this region, the atoms are therefore almost free particles,
interacting with the disordered potential.  The BEC wavefunction
undergoes disorder-induced multiple transmissions and reflections
and is ultimately blocked by a high peak $V(z) > \mu$ of the
speckle potential.  The continually evolving contributions of the
low-density wings of the BEC are responsible for the fluctuations
of $L$ observed in the simulation for $\sigma_V\gtrsim 0.2\mu$,
even once the core of the wavefunction is localized. At
intermediate distances (Fig.~\ref{Fig.3}b), the Thomas-Fermi
approximation is no longer valid, and the density profile exhibits
time-dependent modulations with a length scale intermediate of
$\xi$ and $\Delta z$.

To understand the role of the disordered potential in this
trapping scenario, it is useful to compare this situation with
that of a periodic potential, with a lattice spacing $\Delta z$
and depth $V_{0}=2\sigma_V$. Our model predicts similar behaviour
in the central region, but differs in the low density wings, where
the condensate would continue to expand due to tunnelling between
the lattice sites.  In the lattice, the condensate fragments when
the central density reaches the value $n_0 \backsimeq
V_{0}/g_{1D}$, independent of the lattice spacing. In the case of
the disordered potential, this final density depends on the
statistical distribution  of the optical potential. We can
calculate the probability of finding two speckle peaks of a given
height within a given distance, which when combined with the
condensate expansion dynamics leads to the following estimate for
the final peak density: $n_0\backsimeq 1.25 (\sigma_V/g_{1D})\ln
(0.47 L_{\rm{TF}}/\Delta z)$.  This formula, dependent on both
$\sigma_V$ and $\Delta z$, is in good agreement with our numerical
findings and will be the subject of future experimental
work\cite{nfinal}.

\section{Conclusion}
We have investigated the transport properties of a BEC in a
disordered potential, observing a cross-over from free expansion
to absence of diffusion as the strength of the disorder increases.
Our experimental findings are supported by numerical simulations
and we have discussed a theoretical model which describes a
scenario for disorder-induced trapping.  In the future it would be
interesting to further investigate this highly controllable
system, for example by changing the correlation length of the
disorder or using Bragg spectroscopy to probe the momentum
spectrum of the BEC.

 This work is supported by the CNRS, Minist\`{e}re de la
Recherche, DGA, EU (IST-2001-38863, MRTN-CT-2003-505032),   ESF
(BEC2000+), Marie Curie Fellowship (J.R.), IXSEA-OCEANO (M.H.) and
INTAS (211-855).



\begin{thebibliography}{12}

\bibitem{Clement}D.~Cl\'{e}ment {\it et al.}, to be published in Phys. Rev. Lett.; cond-mat/0506638.

\bibitem{Akkermans}
E.~Akkermans and G.~Montambaux, {\it Physique M\'esoscopique des
\'{E}lectrons et des Photons}, (EDP Science ed., Paris 2004).

\bibitem{anderson}
P.~W.~Anderson, Phys. Rev. {\bf 109}, 5 (1958); Y.~Nagaoka and
H.~Fukuyama (Eds.),  {\it Anderson Localization}, Springer Series
in Solid State Sciences No.~39,
 (Springer, Berlin,
1982); T.~Ando and H.~Fukuyama (Eds.), {\it Anderson
Localization}, Springer Proceedings in  Physics No.~28, (Springer,
Berlin, 1988).

\bibitem{aharony}
A.~Aharony and D.~Stauffer, {\it Introduction to Percolation
Theory} (Taylor \& Francis, London, 1994).

\bibitem{fisher}
M.~P.~A.~Fisher, P.~B.~Weichman, G.~Grinstein, and D.~S.~Fisher,
Phys. Rev. B {\bf 40}, 546 (1989).

\bibitem{parisi}
M.~M\'ezard, G.~Parisi, and M.~A.~Virasoro, {\it Spin Glass and
Beyond} (World Scientific, Singapore, 1987).

\bibitem{damski2003}
B.~Damski {\it et al.}, 
Phys. Rev. Lett. {\bf 91}, 080403 (2003); R.~Roth and K.~Burnett,
J. Opt. B \textbf{5}, S50 (2003); L.~Sanchez-Palencia and
L.~Santos, cond-mat/0502529; A.~Sanpera {\it et al.}, 
Phys. Rev. Lett. {\bf 93}, 040401 (2004).

\bibitem{chips}
see {\it e.g.} J.~Est\`eve {\it et al.},
Phys. Rev. A {\bf } 043629 (2004) and references therein.

\bibitem{Lye}
J.~E.~Lye {\it et al.}, 
cond-mat/0412167;  C.~Fort {\it et al.}, cond-mat/0507144.

\bibitem{Arlt} T.~Schulte {\it et al.}, cond-mat/0507453.

\bibitem{goodman}
J.~W.~Goodman, in {\it Laser Speckle and Related Phenomena},
edited by \linebreak J.~C.~Dainty (Springer-Verlag, Berlin, 1975).

\bibitem{Hug05}
M.~Hugbart {\it et al.}, Eur. Phys. J. D {\bf{35}}, 155 (2005);
cond-mat/0501456.

\bibitem{Thouless}D.~J.~Thouless, Phys. Rev. Lett. {\bf{39}}, 1167
(1977).

\bibitem{orsayBEC}
V. Boyer {\it et al.}, 
Phys. Rev. A {\bf 62}, 021601 (2000).

\bibitem{Pet01}
D.~S.~Petrov, G.~V.~Shlyapnikov, and J.~T.~M.~Walraven, Phys. Rev.
Lett.~{\bf 87}, 050404 (2001); S. Richard {\it et al.}, 
Phys. Rev. Lett. {\bf 91}, 010405 (2003).

\bibitem{scaling}
Yu.~Kagan, E.~L.~Surkov, and G.~V.~Shlyapnikov, Phys. Rev. A~{\bf
54}, R1753 (1996); Y.~Castin and R.~Dum, Phys. Rev. Lett.~{\bf
77}, 5315 (1996).

\bibitem{nfinal} Further details of these properties will be
described in a future publication.
\end{thebibliography}
\end{document}